\documentclass[12pt]{article}
\usepackage{amssymb}
\usepackage{amsmath}
\newcommand{\beq}{\begin{eqnarray}}
\newcommand{\eeq}{\end{eqnarray}}

\newcommand{\al}{\alpha}

\newcommand{\pa}{\partial}

\newcommand{\half}{{\frac 1 2}}
\newcommand{\ihalf}{{\frac i 2}}

\usepackage[top=2.75cm, bottom=2.75cm, left=2.75cm, right=2.75cm]{geometry}

\begin{document}
\thispagestyle{empty}
\begin{flushright} \small
IMPERIAL-TP-2008-CH-01
 \\UUITP-2/08 \\ NORDITA -2008-6\\
\end{flushright}
\smallskip
\begin{center} \LARGE
{\bf    Topological Sigma Models with H-Flux}
 \\[12mm] \normalsize
{\bf C.M.~Hull $^{a,d}$, U.~Lindstr\"om$^{b,c}$, L.~Melo dos Santos$^{d}$,
R. von Unge$^{e}$, and M.~Zabzine$^{b}$} \\[8mm]
 {\small\it
 $^a$The Institute for Mathematical Sciences\\
 Imperial College\\
 53 Prince's Gate, London SW7 2PG, U.K.\\
 ~\\
$^b$Department of Theoretical Physics
Uppsala University, \\ Box 803, SE-751 08 Uppsala, Sweden \\
~\\
$^c$NORDITA, Roslagstullsbacken 23\\
SE-10691 Stockholm Sweden\\
~\\
$^d$The Blackett Laboratory, Imperial College London\\
Prince Consort Road, London SW7 2AZ, U.K.\\
~\\
$^{e}$Institute for Theoretical Physics, Masaryk University, \\ 61137 Brno, Czech Republic
\\~\\
}
\end{center}
\vspace{10mm}
\centerline{\bfseries Abstract} \bigskip

\noindent  We investigate the topological theory obtained by twisting the
  ${\cal{N}}=(2,2)$ supersymmetric nonlinear sigma model with target a bihermitian space
with torsion. For the special case in which the two complex
structures commute, we show that  the action is a $Q$-exact term
plus a quasi-topological term. The quasi-topological term is locally
given by a closed two-form which corresponds to a flat
gerbe-connection and generalises the usual topological term of the
A-model. Exponentiating it gives a Wilson surface,
 which can be regarded as a generalization of a Wilson line. This makes the quantum theory globally well-defined.
\eject
\normalsize

\section{Introduction}
The topological sigma model was introduced in \cite{Witten:1988xj} for
a target which is a symplectic manifold. In \cite{Witten:1991zz,
Labastida:1991qq} it was observed that for the K\"ahler case
the topological  sigma model  can be obtained by twisting the ${\cal
N}=(2,2)$ sigma model with K\"ahler target, with a certain linear
combination of the four supercharges becoming a scalar charge $Q$,
sometimes referred to as a BRST operator. The   action  
is the sum of a $Q$-exact term and a topological term, so that the
path integral is given as a sum over topological sectors weighted by the
exponential of the  topological term.
 Moreover the path integral of the twisted model is localized on the fixed points of the
 $Q$-action. For the A-model of \cite{Witten:1991zz}, the theory localises on holomorphic
 maps and the topological term is the pullback of the K\"ahler form.
For a comprehensive review and   applications to string theory see, e.g., \cite{Clay}.

The most general  ${\cal N}=(2,2)$ sigma model  has both a kinetic
term given by a metric $g$ and a Wess-Zumino term defined by a
closed 3-form $H$, so that in each patch $U_\alpha$ there is a 2-form $B_\alpha$ with $H=dB_\alpha$ \cite{Gates:1984nk}, \cite{Howe:1985pm}. The geometry of the target
space is bihermitian, i.e., the   metric $g$ is hermitian with
respect to two complex structures $J_\pm$, and these are covariantly
constant with respect to the covariant derivatives with torsion
\beq\label{covarGKG}
\nabla^{(\pm)}J_\pm=0~,~~~~~~~\nabla^{(\pm)} = \nabla \pm \frac{1}{2}~ g^{-1} H~,
\eeq
 where $\nabla$ is the Levi-Civita connection.  Alternatively,  the conditions (\ref{covarGKG})
  can be rewritten as the following  integrability conditions
 \beq\label{integrabilityforGKS} 
  H = d_+^c \omega_+ = - d_-^c \omega_-~,
 \eeq
  where $\omega_\pm =gJ_\pm$ and $d_\pm^c$ are the $i(\bar{\partial}-\partial)$ operators  for 
    the complex structures $J_\pm$.     
     When the complex structures commute, the geometry also carries a local product structure. We refer to this special case as a bihermitian Local Product (BiLP) geometry. It was shown in \cite{Gates:1984nk}
   that this case has a manifest ${\cal N}=(2,2)$ superspace description in terms of chiral and twisted chiral superfields. Bihermitian geometry has been given an alternative formulation as
Generalized K\"ahler Geometry \cite{gualtieri} and subsequently in  \cite{Lindstrom:2005zr} it
has been demonstrated that this geometry is locally described in
${\cal N}=(2,2)$ superspace using semi-(anti) chiral superfields \cite{Buscher:1987uw} in
addition to chiral and twisted ones.

  The only term explicitly  dependent on $B$ is the the WZ-term, which is proportional to
$$ \int  X^*(B)= \frac 1 2 \int d^2\xi~B_{AB} \epsilon ^{\mu \nu }\partial _\mu  X^A \partial _\nu X^B~.$$
The terms involving fermions depend on  $B$ only through the field strength $H$. In Euclidean signature,  Wick-rotation leads   to an imaginary
       WZ term with a factor of 
        \lq i'   in front of $B$.

    For the quantum theory to be well-defined, it is necessary that $H \in H^3(M, \mathbb{Z})$, so that there is   a gerbe
   with connection $\{ B_\alpha\}$ whose curvature is $H$.
   For the path integral, if
   $H_2(M)$ is trivial, 
  then  the image $c_2=X(\Sigma_2)$ of a compact world-sheet $(\Sigma_2$  is the boundary of  a  three dimensional submanifold $c_3$) and the WZ term can be written as an integral of $H$ over $c_3$, so that  the WZ-term
  depends on $H$ only.
  If
   $H_2(M)$ is non-trivial, 
 it is not sufficient to specify $H$, and a choice of $B$ must be specified. Then  the exponent of the WZ-term is defined as the holonomy of a gerbe over
     $X(\Sigma_2)$, see,  e.g., formula  in the Appendix 
     (for further details  on gerbes and gerbe holonomy the reader may consult \cite{Belov:2007qj, hitchingerbe, chatterjee, mackaay}). It is important
      to stress that  to define a gerbe holonomy we in general need full information
       about the gerbe connection $\{ B_\alpha \}$, not just $H$ alone.

The twisting of the general ${\cal N}=(2,2)$ sigma model   with torsion was considered in
  \cite{Kapustin:2004gv} and discussed further in \cite{Bredthauer:2006hf},
     \cite{Zucchini:2006ii}, \cite{Chuang:2006vt}.
However, there were problems in writing the action as the sum of a $Q$-exact term and a topological term, so that it was hard to understand the structure of the path integral as a weighted sum.
Here we shall  write the action in just such a form, in the special case in which the  complex structures commute, so that the geometry is BiLP.
The $Q$-exact term  $QV$ can be found from the ${\cal N}=(2,2)$ superspace formulation, in which the action is given by the superspace integral of a potential $K$ depending on all superfields.
Grassmann coordinates $\theta, \theta ^1 , \theta ^2 , \theta ^3$ can be chosen such that
$Q= \partial/\partial \theta$, and hence $V$ is  given by integrating $K$ over $  \theta ^1 , \theta ^2 , \theta ^3$.
Strictly speaking, it is determined in this way up to a total derivative term, as the usual superspace approach is not sufficiently careful with boundary terms.
 Although we believe that twisting  can nevertheless be  performed in superspace \cite{collab}, we choose to circumvent the derivative question and related issues by using a component presentation
in which the total derivative terms are fixed.

The local product structure splits the coordinates locally into two sets, $\phi ^i$ and $\chi ^a$.
One set of coordinates are the leading components of chiral superfields and the other set are the leading components of twisted chiral superfields;  which can be interchanged by a coordinate redefinition in superspace.
The A-twist of the model in which the $\phi ^i$ are chiral  and the $\chi ^a$ are  twisted chiral is the same as the B-twist of the model in which the $\phi ^i$ are twisted  chiral  and the $\chi ^a$ are  chiral, so
 all cases are covered by considering, say,  the B-twist of the  general model with arbitrary numbers of chiral superfields and twisted chiral superfields\footnote{Note that T-duality and mirror symmetry interchanges A and B-twists, and thus chiral and twisted chiral fields.}.
We shall discuss the B-twist here.

The main result of the paper can be summarized as follows.
 The action of the twisted model can be written as a sum
 of $Q$-exact term and a \lq quasi-topological term'.
 This reduces to the usual topological term of the  topological sigma-model when the target space is Kahler, but with a non-trivial $B$ field, this term in the action is not well-defined.
 However, its exponential is well-defined and gives the holonomy of a flat gerbe, so that the quantum theory is well-defined, with the path integral weighted by these holonomies.
 
  The structure of the paper is as follows.  In section \ref{background} we present some
  background information about the superspace formalism and topological twist.
  In section \ref{twisted} the component analysis is done. The twisted action is written
   as a sum of a $Q$-exact term and the pull-back of a locally defined closed form.
 Section \ref{gerbes}  is devoted to the geometrical interpretation of the quasi-topological
   term using the language of flat gerbes.
 In section \ref{end} some comments and speculations are presented. In Appendix 
  we briefly review
  the definition of holonomy for line bundles and gerbes.

\section{Background}
\label{background}

In this section we define the twist from the point of view of
superspace.

The original ${\cal N}=(2,2)$ sigma model with Minkowski signature
has the Lorentz group $SO(1,1)$ acting on the world sheet
coordinates.
In addition, there is an   $SO(2)\times SO(2)$ R-symmetry acting on the  superspace Grassman variables
($\theta^{1+}$,$\theta^{2+}$,$\theta^{1-}$,$\theta^{2-}$),
  with one $SO(2)$ acting on the positive chirality
  odd variables $\theta^{I +}$
  and the other acting on the negative chirality ones $\theta^{I -}$ ($I =1,2$).
   It is important to remember that $\theta^{I+}$ and
$\theta^{I-}$ transform as Majorana-Weyl spinors, i.e. each is a
one-component real spinor.
 The symmetry group of the sigma model with
Minkowski signature is then
\begin{equation}
SO(1,1)\times SO(2)\times SO(2)~.
\end{equation}
The $R$-rotations  act on superfields $\Phi$ as vector or axial rotations:
\beq
V&:& \Phi^i(x,\theta^{\pm}, \bar \theta^{\pm}) \mapsto e^{i\al q^i_V}\Phi^i(x,e^{-i\al}\theta^{\pm}, e^{i\al}\bar \theta^{\pm})\cr
A&:&\Phi^i(x,\theta^{\pm}, \bar \theta^{\pm}) \mapsto e^{i\beta q^i_A}\Phi^i(x,e^{\mp i\beta}\theta^{\pm}, e^{\pm i\beta}\bar \theta^{\pm})~,
\eeq
 where $q_V$ and $q_A$ are the vector and axial $R$-charges respectively.
 Here $\theta^{\pm} = \frac{1}{\sqrt{2}}(\theta^{1\pm}+i \theta^{2\pm})$. 
To  twist the model  as in  \cite{Witten:1991zz,
Labastida:1991qq}, 
one must first  Wick rotate, so that the Lorentz group becomes
SO(2). 

In Euclidean signature, we want to treat fields and their complex conjugates as formally independent, in order to allow e.g. a B-twist in which positive and negative chirality fields are twisted differently.
This means that we need to consider  the {\it complexification}
   of the Euclidean theory (in analogy with CFT) and as a  result consider the complexification
    of the ${\cal N}=(2,2)$ superspace.  
We will say more about this complexification elsewhere \cite{newcool}, but here simply note that
the complexified ${\cal N}=(2,2)$
Euclidean  sigma model has the symmetry group (Lorentz and R symmetries)
\begin{equation}
SO(2,\mathbb{C})\times SO(2,\mathbb{C})\times SO(2,\mathbb{C})~,
\end{equation}
allowing the possibility of A and B twists. In the
complexified
 superspace we treat  the (twisted) chiral and (twisted) anti-chiral as independent fields.

The ${\cal N}=(2,2)$ Euclidean supersymmetry algebra\footnote{In the algebra (\ref{N=22Eucl}) it would
 be natural to remove the imaginary \lq $i$' from the right hand side. However we prefer to keep it in order
  to preserve the formal similarities with   Minkowski-signature superspace.} is
\beq\label{N=22Eucl}
 \left\{\mathbb{D}_+,\bar{\mathbb{D}}_+\right\}
&=& i\partial~,\cr \left\{\mathbb{D}_-,\bar{\mathbb{D}}_-\right\} 
&=&
i\bar\partial~, \eeq where $\partial \equiv \pa /\pa z$ and $z=x^1+
i x^2$. As a result of twisting, two of the supercharges become
scalars $Q$, $\hat{D}$, and two become vectors $D_z$, $D_{\bar z}$.
For the A and B twists, \beq\label{newops} &A-twist:& \quad D_z =
\bar{\mathbb{D}}_-,\;\; D_{\bar z} = \mathbb{D}_+,\;\; Q =
\frac12\left(\bar{\mathbb{D}}_+ + \mathbb{D}_-\right),\;\; \hat{D} =
\frac{1}{2i}\left(\bar{\mathbb{D}}_+ - \mathbb{D}_-\right)~,\cr
&B-twist:& \quad D_z = {\mathbb{D}}_-,\;\; D_{\bar z} =
\mathbb{D}_+,\;\; Q = \frac12\left(\bar{\mathbb{D}}_+ +
\bar{\mathbb{D}}_-\right),\;\; \hat{D} =
\frac{1}{2i}\left(\bar{\mathbb{D}}_+ - \bar{\mathbb{D}}_-\right)~.
\eeq

The new operators obey the algebra
\beq
\{Q,D_z\}= \ihalf\bar\partial,\;\;
\{Q,D_{\bar z} \} = \ihalf\partial,\;\;
\{\hat{D},D_z\} = -\half\bar\partial,\;\;
\{\hat{D},D_{\bar z}\} = \half\partial~,
\eeq
for both twists. In fact, from the superspace point of view  only one twist exists, the $A$ and $B$ twists being related by a coordinate transformation exchanging $\theta^-$ with $\bar\theta^-$, or equivalently, by exchanging the  chiral and twisted chiral fields. For concreteness, from now on we shall focus on the $B$-twist.

In Minkowski superspace, chiral ($\Phi$) and twisted chiral fields
($\chi$) obey the constraints \beq \bar{\mathbb{D}}_\pm\Phi=0&,&
\bar{\mathbb{D}}_+\chi={\mathbb{D}}_-\chi=0~,\cr
\mathbb{D}_\pm\bar{\Phi}=0&,&\mathbb{D}_+\bar{\chi}=\bar{\mathbb{D}}_-\bar{\chi}=0~,
\label{constraints} \eeq along with their conjugates. In the 
Euclidean theory we need to consider $\phi$, $\bar{\phi}$, $\chi$,
$\bar{\chi}$ as independent with  constraints
(\ref{constraints}).

 For a
$B$-twist, the chiral constraints  on the superfields $\Phi$, $\bar{\Phi}$
may be re-expressed in terms of the new operators as
\beq\label{chiral} Q\Phi = 0,\;\; \hat{D}\Phi = 0,\;\; D_z \bar\Phi
= 0,\;\; D_{\bar z} \bar\Phi = 0 ~, \eeq while the twisted chiral
superfields $\chi$, $\bar{\chi}$ obey \beq\label{twist} D_z \chi =
0,\;\; Q\chi = -i\hat{D}\chi,\;\; D_{\bar z} \bar\chi = 0,\;\;
Q\bar\chi = i\hat{D}\bar\chi~. \eeq Denoting the
$\theta$-independent part by a vertical bar, we define the standard
components of our superfields as \beq\label{scomps} \Phi &:
&\phi\equiv \Phi |,\quad \psi_\pm \equiv {\mathbb{D}}_\pm\Phi
|,\quad F\equiv {\mathbb{D}}_+{\mathbb{D}}_-\Phi |~,\cr \chi &:&
\chi\equiv\chi |,\quad \lambda_+\equiv {\mathbb{D}}_+\chi |,\quad
\lambda_- \equiv \bar{\mathbb{D}}_-\chi |,\quad G\equiv
{\mathbb{D}}_+ \bar{\mathbb{D}}_-\chi |~, \eeq along with similar expressions for
$\bar{\Phi}$ and $\bar{\chi}$.
 For completeness we also define the components with
respect to the new operators (\ref{newops}) and give their relation
to the components (\ref{scomps})
\beq \chi &:& \tilde\chi = \chi|
=\chi,\;\; \rho_{\bar z} = D_{\bar z}\chi|=\lambda_+,\;\; \eta =
-i\hat D \chi|,=\half \lambda_-\;\; G_{\bar z} = iD_{\bar z} \hat D
\chi |=-\half G\cr \bar\chi &:& \bar{\tilde\chi} = \bar\chi
|,=\bar\chi\;\; \bar\rho_{ z} = D_{ z}\bar\chi|,=\bar\lambda_-\;\;
\bar\eta = i\hat D\bar\chi|=\half\bar\lambda_+,\;\; \bar G_{z} =
-iD_{z}\hat D \bar\chi |=-\half\bar G,\cr \Phi &:& \tilde\phi = \Phi
|=\phi,\;\; \psi_z = D_z \Phi |=\psi_-,\;\; \psi_{\bar z} = D_{\bar
z} \Phi |=\psi_+,\;\; F_{z\bar z} = D_z D_{\bar z}\Phi|=F,\cr
\bar\Phi&:&\bar{\tilde\phi} = \bar\Phi |=\bar\phi,\;\; \bar\varphi =
Q\bar\Phi |=\half(\bar\psi_++\bar\psi_-),\;\; \bar\zeta = \hat D
\bar\Phi |=-\ihalf (\bar\psi_+-\bar\psi_-)~,\cr &&\bar H = Q\hat D
\bar\Phi |=-\ihalf\bar F~. \eeq We choose to express the $BRST$
transformations (generated by the charge $Q$) in terms of the
non-standard components: \beq \chi &:& \delta \chi = \eta,\;\;\delta
\eta = 0,\;\; \delta \rho_{\bar z} = G_{\bar z} + \ihalf\partial
\chi,\;\; \delta G_{\bar z} = -\ihalf\partial\eta~,\cr \bar\chi &:&
\delta \bar \chi =\bar\eta,\;\; \delta \bar\eta = 0,\;\; \delta
\bar\rho_{z} = \bar G_{z} + \ihalf\bar\partial\bar\chi,\;\; \delta
\bar G_{z} =-\ihalf\bar\partial\bar\eta,\cr \Phi &:& \delta \tilde
\phi =0,\;\; \delta \psi_z = \ihalf\bar{\partial} \tilde\phi,\;\; \delta
\psi_{\bar z} = \ihalf\partial\tilde\phi,\;\; \delta F_{z\bar z}
=- \ihalf(\partial\psi_{ z}-\bar\partial\psi_{\bar z}),\cr \bar\Phi
&:& \delta \bar{\tilde\phi} = \bar\varphi,\;\;\delta \bar\varphi =
0,\;\; \delta\bar\zeta = \bar H,\;\; \delta \bar H = 0. \eeq

We now have the option of continuing our analysis in superspace
where the $Q$ exact term in the Lagrangian can  be derived once we
have decided which set of components to use. This line of  attack
will be followed elsewhere \cite{collab}.
Alternatively we may turn directly to a component treatment which is
what we do in the next section.
If we do not use complex conjugation in our calculations
 then the formal manipulations in Minkowski and complexified Euclidean space or  superspace are exactly the same.

\section{The twisted model}
\label{twisted}

The approach taken in this section is   to start
from a specific form of the component Lagrangian and then carefully
keep track of all total derivative contributions under variations. The component
approach also allows us to keep the considerations general enough to
allow, e.g., for almost complex geometries.

The supersymmetry transformations we need follow from the superspace
transformation rule
 \beq\label{susyq} \delta \Sigma
=[\al_-Q_++\al_+Q_- +\tilde \al_-\bar Q_++\tilde\al_+\bar Q_-,\Sigma]~,
\eeq for any superfield $ \Sigma$. Since we will be interested in
the transformations  of the components arrived at by taking
$\theta$-independent parts of various superfields, we may replace
the super-charges in (\ref{susyq}) by covariant derivatives according
to : \beq\label{susyd} \delta \Sigma
|=i[\al_-{\mathbb{D}}_++\al_+{\mathbb{D}}_-+\tilde\al_-\bar
{\mathbb{D}}_+ +\tilde\al_+\bar
{\mathbb{D}}_-,\Sigma]|~,\label{trans} \eeq where the vertical bar
denotes the  $\theta$-independent part. Using (\ref{trans}) for the
superfield $\Sigma$ given by $\Phi$, $D\Phi$, $D^2\Phi$, gives the
transformations for the chiral multiplet used in
\cite{Witten:1991zz},
\begin{eqnarray}\label{chtrans}
\delta\phi^i&=&i\alpha_-\psi^i_++i\alpha_+\psi_-^i~,\nonumber\\
\delta\phi^{\bar{i}}&=&i\tilde{\alpha}_-\psi_+^{\bar{i}}+i\tilde{\alpha}_+\psi_-^{\bar{i}}~,\nonumber\\
\delta\psi_+^i&=&-\tilde{\alpha}_-\partial\phi^i-i\alpha_+F^i~,\nonumber\\
\delta\psi_+^{\bar{i}}&=&-\alpha_-\partial\phi^{\bar{i}}+i\tilde{\alpha}_+F^{\bar{i}}~,\\
\delta\psi_-^i&=&-\tilde{\alpha}_+\bar\partial\phi^i+i\alpha_-F^i~,\nonumber\\
\delta\psi_-^{\bar{i}}&=&-\alpha_+\bar\partial\phi^{\bar{i}}-i\tilde{\alpha}_-F^{\bar{i}}~,\nonumber\\
\delta
F^i&=&-\tilde{\alpha}_-\partial\psi_-^i+\tilde{\alpha}_+\bar\partial\psi_+^i~,\nonumber\\
\delta
F^{\bar{i}}&=&\alpha_-\partial\psi_-^{\bar{i}}-\alpha_+\bar\partial\psi_+^{\bar{i}}~,\nonumber
\label{chiral transformations}
\end{eqnarray}
where we have introduced a set of chiral multiplets labeled by
$i,j,...$ and a set of  antichiral multiplets labelled by $\bar{i}$,
$\bar{j}$, ... The
 transformations of the components of twisted multiplets labelled by
 $a,b,...$ and anti twisted multiplets labelled by $\bar{a}$,
 $\bar{b}$, ...
are found  similarly to be
\begin{eqnarray}\label{twtrans}
\delta\chi^a&=&i\alpha_-\lambda_+^a+i\tilde{\alpha}_+\lambda_-^a~,\nonumber\\
\delta\chi^{\bar{a}}&=&i\alpha_+\lambda_-^{\bar{a}}+i\tilde{\alpha}_-\lambda_+^{\bar{a}}~,\nonumber\\
\delta\lambda_+^a&=&-\tilde{\alpha}_-\partial\chi^a-i\tilde{\alpha}_+G^a~,\nonumber\\
\delta\lambda_+^{\bar{a}}&=&-\alpha_-\partial\chi^{\bar{a}}+i\alpha_+G^{\bar{a}}~,\\
\delta\lambda_-^a&=&-\alpha_+\bar\partial\chi^a+i\alpha_-G^a~,\nonumber\\
\delta\lambda_-^{\bar{a}}&=&-\tilde{\alpha}_+\bar\partial\chi^{\bar{a}}-i\tilde{\alpha}_-G^{\bar{a}}~,\nonumber\\
\delta
G^a&=&-\tilde{\alpha}_-\partial\lambda_-^a+\alpha_+\bar\partial\lambda_+^a~,\nonumber\\
\delta
G^{\bar{a}}&=&\alpha_-\partial\lambda_-^{\bar{a}}-\tilde{\alpha}_+\bar\partial\lambda_+^{\bar{a}}\nonumber~.
\label{twisted chiral transformations}
\end{eqnarray}

As shown in \cite{Gates:1984nk},  the ${\cal N}=(2,2)$ Lagrangian
used in \cite{Witten:1991zz} can be generalized to include the
twisted chiral fields. It then reads
 \beq\label{fullcompaction} S =
\int d^2\xi\left( E_{AB} \partial X^A \bar\partial X^B +\frac12
g_{AB}\psi^A_+ i\nabla^{(+)}\psi_+^B +\frac12 g_{AB}\psi^A_-
i\bar\nabla^{(-)}\psi^B_- +\frac14 R^{(+)}_{ABCD} \psi_+^A \psi_+^B
\psi_-^C \psi_-^D \right), \eeq where $A,B,...$ label the
(anti)chiral and twisted (anti)chiral fields and the connection used
in the covariant derivative and curvature have torsion
$\Gamma_{ABC}^{(\pm)}\equiv\Gamma_{ABC}^{(0)} \pm \frac{1}{2}H_{ABC}$ with
$\Gamma^{(0)}$ the Levi-Civita connection and 
\beq H\equiv dB
~,\quad
E_{AB}\equiv g_{AB}+ B_{AB}~. \eeq 
The explicit form of the action may
be found using the components defined in (\ref{scomps}).

 The  equations of motion for the auxiliary fields $F$ and $G$ are,
\beq
&&K_{\bar{j}i}F^i=K_{\bar{j}kl}\psi_-^l\psi_+^k+K_{\bar{j}a\bar{b}}\lambda_-^{\bar{b}}\lambda_+^a+K_{\bar{j}k\bar{a}}\lambda_-^{\bar{a}}\psi_+^k+K_{\bar{j}ak}\psi_-^k\lambda_+^a\cr
&&K_{\bar{b}a}G^a=K_{\bar{b}cd}\lambda_-^d\lambda_+^c+K_{\bar{b}i\bar{j}}\psi_-^{\bar{j}}\psi_+^i+K_{\bar{b}c\bar{i}}\psi_-^{\bar{i}}\lambda_+^c+K_{\bar{b}ic}\lambda_-^c\psi_+^i,
\eeq
and may be used to go partially on-shell and eliminate the auxiliary fields in (\ref{chtrans}) and (\ref{twtrans}).

To construct a topological model out of this Lagrangian, we use the
same procedure as  in \cite{Witten:1991zz}, and  twist the Lorentz group with either  a vector or axial subgroup of the R-symmetry group,
so that two of the supercharges become scalars.

Using the R-symmetries available in the (2,2)-algebra we twist the
Lorentz transformations of the fields, such that for the B-model
\begin{eqnarray}
Q_B&=&\bar{Q}_++\bar{Q}_-~,\\\nonumber
Q^T&=&\bar{Q}_+-\bar{Q}_-~,
\end{eqnarray}
are scalar charges.

As in \cite{Witten:1991zz},  we expect to obtain a topological model
after the twisting with a Lagrangian $L_{total}$ written 
\begin{equation}
L_{total}=L_B+L_{top} ~.\label{structure of the action}
\end{equation}
The first term $L_B$ would be a $Q_B$ exact term ($L_B=Q_BV'$). The
second term $L_{top}$ is expected to be some kind of  \lq topological term'  that is a total derivative and does not affect the equations
of motion. If this term is given by a closed 2-form, then its
integral is a topological invariant depending on the cohomology
class of the 2-form and the homology class of the embedding of the world-sheet in the target. 
 In the case involving only
 chiral and antichiral fields, for the A-model, this term turns out to be the K\"ahler
 form $\omega$ of the target space and its integral is the degree of the holomorphic map, \cite{Witten:1991zz}. To find the total derivative term here, we will first calculate the exact part
 $L_B$ of the Lagrangian and subtract it from the full
 Lagrangian $L_{total}$. Here we find it not to be given by a globally
 defined 2-form, but instead by a flat gerbe connection, so it does
 not determine a cohomology class but instead leads to a generalisation
 of a Wilson line.

 Then $V'$ is calculated by  taking a potential  
 depending on all the fields,
 $K(\phi,\bar{\phi},\chi,\bar{\chi})$, and acting on it with the other 3 supersymmetries,
 \begin{equation}
 V'=Q^TQ_-Q_+K~.
 \end{equation}
(We note in passing  that $L_B$ is proportional to the full
superspace integral of the superspace Lagrangian
 $K(\Phi,\bar\Phi,\chi,\bar\chi)$  up to surface terms which depend
   on the  precise prescription for the superspace measure.)
 We can use the transformations (\ref{chiral transformations}) and
 (\ref{twisted chiral transformations}) to perform the first two
 steps, and use the transformations for the scalar supercharges,
 \begin{eqnarray}
\delta\phi^{\bar{i}}&=&i\alpha_B(\psi_+^{\bar{i}}+\psi_-^{\bar{i}})+i\alpha^T(\psi_-^{\bar{i}}-\psi_+^{\bar{i}})~,\nonumber\\
\delta\psi_+^i&=&-\alpha_B\partial\phi^i+\alpha^T\partial\phi^i~,\nonumber\\
\delta\psi_-^i&=&-\alpha_B\bar\partial\phi^i-\alpha^T\bar\partial\phi^i~,\nonumber\\
\delta
F^i&=&-\alpha_B(\partial\psi_-^i-\bar\partial\psi_+^i)+\alpha^T(\partial\psi_-^i+\bar\partial\psi_+^i)\\
\delta\chi^a&=&i\alpha_B\lambda_-^a+i\alpha^T\lambda_-^a~,\nonumber\\
\delta\chi^{\bar{a}}&=&i\alpha_B\lambda_+^{\bar{a}}-i\alpha^T\lambda_+^{\bar{a}}~,\nonumber\\
\delta\lambda_+^a&=&\alpha_B(-\partial\chi^a-iG^a)+\alpha^T(\partial\chi^a-iG^a)~,\nonumber\\
\delta\lambda_-^{\bar{a}}&=&\alpha_B(-\bar\partial\chi^{\bar{a}}-iG^{\bar{a}})+\alpha^T(-\bar\partial\chi^{\bar{a}}+iG^{\bar{a}})~,\nonumber
\end{eqnarray}
for the last variation of the potential $K$.

Performing   these variations, we find the explicit form of
$V'$,
\begin{eqnarray}
V'&=&-iK_{i\bar{j}}\psi_-^{\bar{j}}F^i+iK_{i\bar{j}}\psi_+^{\bar{j}}F^i-iK_{ia}\lambda_-^aF^i+iK_{i\bar{a}}\lambda_+^{\bar{a}}F^i\nonumber\\
&&-K_i\partial\psi_-^i-K_i\bar\partial\psi_+^i-iK_{ij\bar{k}}\psi_-^{\bar{k}}\psi_+^j\psi_-^i+iK_{ij\bar{k}}\psi_+^{\bar{k}}\psi_+^j\psi_-^i\nonumber\\
&&-iK_{ija}\lambda_-^a\psi_+^j\psi_-^i+iK_{ij\bar{a}}\lambda_+^{\bar{a}}\psi_+^j\psi_-^i-K_{ij}\partial\phi^j\psi_-^i-K_{ij}\psi_+^j\bar\partial\phi^i\nonumber\\
&&-iK_{\bar{a}b\bar{i}}\psi_-^{\bar{i}}\lambda_+^b\lambda_-^{\bar{a}}+iK_{\bar{a}b\bar{i}}\psi_+^{\bar{i}}\lambda_+^b\lambda_-^{\bar{a}}-iK_{\bar{a}bc}\lambda_-^c\lambda_+^b\lambda_-^{\bar{a}}+iK_{\bar{a}b\bar{c}}\lambda_+^{\bar{c}}\lambda_+^b\lambda_-^{\bar{a}}\nonumber\\
&&-K_{\bar{a}b}\partial\chi^b\lambda_-^{\bar{a}}+iK_{\bar{a}b}G^b\lambda_-^{\bar{a}}-K_{\bar{a}b}\lambda_+^b\bar\partial\chi^{\bar{a}}+iK_{\bar{a}b}\lambda_+^bG^{\bar{a}}\nonumber\\
&&-iK_{ia\bar{j}}\psi_-^{\bar{j}}\lambda_+^a\psi_-^i+iK_{ia\bar{j}}\psi_+^{\bar{j}}\lambda_+^a\psi_-^i-iK_{iab}\lambda_-^b\lambda_+^a\psi_-^i+iK_{ia\bar{b}}\lambda_+^{\bar{b}}\lambda_+^a\psi_-^i\nonumber\\
&&-K_{ia}\partial\chi^a\psi_-^i+iK_{ia}G^a\psi_-^i-K_{ia}\lambda_+^a\bar\partial\phi^i\nonumber\\
&&-iK_{\bar{a}i\bar{j}}\psi_-^{\bar{j}}\psi_+^i\lambda_-^{\bar{a}}+iK_{\bar{a}i\bar{j}}\psi_+^{\bar{j}}\psi_+^i\lambda_-^{\bar{a}}-iK_{\bar{a}ib}\lambda_-^b\psi_+^i\lambda_-^{\bar{a}}+iK_{\bar{a}i\bar{b}}\lambda_+^{\bar{b}}\psi_+^i\lambda_-^{\bar{a}}\nonumber\\
&&-K_{\bar{a}i}\partial\phi^i\lambda_-^{\bar{a}}-K_{\bar{a}i}\psi_+^i\bar\partial\chi^{\bar{a}}+iK_{\bar{a}i}\psi_+^iG^{\bar{a}}~.
\end{eqnarray}
Here indices on $K$ denote partial derivatives, so that e.g.
$$K_{i\bar{j}}
= \partial_i \partial_{\bar{j}} K~.
$$
To calculate the exact term, we still have to  $BRST$-transform
$V'$. Just as in the chiral case, the topological term will be
associated with the part of the action involving only the scalars. For $Q_BV'$ the
purely bosonic part is given by 
\begin{eqnarray}
\nonumber Q_BV'&=&K_{i\bar{j}}\partial\phi^{\bar{j}}\bar\partial\phi^i+K_{i\bar{j}}\bar\partial\phi^{\bar{j}}\partial\phi^i\\
&&-K_{\bar{a}b}\partial\chi^b\bar\partial\chi^{\bar{a}}-K_{\bar{a}b}\partial\chi^b\bar\partial\chi^{\bar{a}}\nonumber\\
&&+K_{i\bar{a}}\partial\chi^{\bar{a}}\bar\partial\phi^i+K_{ia}\bar\partial\chi^a\partial\phi^i\nonumber\\
&&-K_{ia}\partial\chi^a\bar\partial\phi^i-K_{\bar{a}i}\partial\phi^i\bar\partial\chi^{\bar{a}}+
{\rm fermi\, terms }~. \label{exact}
\end{eqnarray}
Note that the fixed points of the $Q$-transformations  are given by holomorphic twisted chiral maps and by constant chiral ones. Thus   arguments   show that  the
 theory is localized on those maps.

Let us now focus on the full Lagrangian of the system. We know from \cite{Gates:1984nk} that the target
space   is a bihermitian manifold. We choose a local
chart for that manifold such that
\begin{eqnarray}
{J_{\pm}}^i_{\phantom{i}j}&=&i\delta^i_{\phantom{i}j}~,\nonumber\\
{J_{\pm}}^{\bar{i}}_{\phantom{i}\bar{j}}&=&-i\delta^{\bar{i}}_{\phantom{i}\bar{j}}~,\nonumber
\end{eqnarray}
\begin{eqnarray}
{J_+}^a_{\phantom{a}b}=i\delta^a_{\phantom{a}b}~,&
&{J_-}^a_{\phantom{a}b}=-i\delta^a_{\phantom{a}b}~,\nonumber\\
{J_+}^{\bar{a}}_{\phantom{a}\bar{b}}=-i\delta^{\bar{a}}_{\phantom{a}\bar{b}}~,&
&{J_-}^{\bar{a}}_{\phantom{a}\bar{b}}=i\delta^{\bar{a}}_{\phantom{a}\bar{b}}~,
\end{eqnarray}
where $J_\pm$ are the complex structures of the manifold. The metric
$g$ is 
\begin{eqnarray}
g_{i\bar{j}}=K_{i\bar{j}}& &g_{a\bar{b}}=-K_{a\bar{b}}~.
\end{eqnarray}
We  then define the two-forms $\omega_\pm=gJ_\pm$, which 
 in this coordinate system are
\begin{equation}
\omega_\pm=-iK_{i\bar{j}}d\phi^i\wedge d\phi^{\bar{j}}\pm
iK_{a\bar{b}}d\chi^a \wedge d\chi^{\bar{b}}~.
\end{equation}
The full Lagrangian $L_{total}$ has two distinct geometrical parts,
terms depending on the metric $g$ of the target space and terms
depending on a $B$-field on the target space. For the $B$-field,
a useful gauge is the one of \cite{Gates:1984nk}.
 In this gauge, $B$ is chosen to be $B_-$, where $d B_-=H$ and 
   the $(1,1)$ component of $B_-$ with respect to $J_-$ vanishes. 
    Then with $B=B_-$,   the bosonic part of $L_{total}$ reads
\begin{equation}
L_{total}=K_{i\bar{j}}\partial\phi^i\bar\partial\phi^{\bar{j}}+K_{i\bar{j}}\bar\partial\phi^i\partial\phi^{\bar{j}}-K_{a\bar{b}}\partial\chi^a\bar\partial\chi^{\bar{b}}-K_{a\bar{b}}\bar\partial\chi^a\partial\chi^{\bar{b}}-K_{i\bar{a}}d\phi^i\wedge
d\chi^{\bar{a}}-K_{\bar{i}a}d\phi^{\bar{i}}\wedge d\chi^a~.
\label{local action}
\end{equation}
We can write
$L_{total}$ explicitly as 
\begin{eqnarray}
L_{total}&=&K_{i\bar{j}}\partial\phi^i\bar\partial\phi^{\bar{j}}+K_{i\bar{j}}\bar\partial\phi^i\partial\phi^{\bar{j}}\nonumber\\
&&-K_{a\bar{b}}\partial\chi^a\bar\partial\chi^{\bar{b}}-K_{a\bar{b}}\bar\partial\chi^a\partial\chi^{\bar{b}}\nonumber\\
&&-K_{i\bar{a}}\partial\phi^i\bar\partial\chi^{\bar{a}}-K_{\bar{i}a}\partial\phi^{\bar{i}}\bar\partial\chi^a\nonumber\\
&&+K_{a\bar{i}}\partial\chi^a\bar\partial\phi^{\bar{i}}+K_{\bar{a}i}\partial\chi^{\bar{a}}\bar\partial\phi^i~.
\label{total}
\end{eqnarray}
We are now ready to determine the total derivative term of the
action $L_{top}$. Subtracting the exact part of the action $Q_BV'$, given in
(\ref{exact}),  from $L_{total}$, given in (\ref{total}), we get
\begin{eqnarray}
L_{top}&=&K_{a\bar{b}}\partial\chi^a\bar\partial\chi^{\bar{b}}-K_{a\bar{b}}\bar\partial\chi^a\partial\chi^{\bar{b}}\nonumber\\
&&+K_{ai}\partial\chi^a\bar\partial\phi^i-K_{ai}\bar\partial\chi^a\partial\phi^i\nonumber\\
&&+ K_{a\bar{i}}\partial\chi^a\bar\partial\phi^{\bar{i}}-K_{a\bar{i}}\bar\partial\chi^a\partial\phi^{\bar{i}}~.
\end{eqnarray}
Focusing on the target space, the term can be written as a pullback of a two-form,
\begin{equation}
L_{top}= X^*(K_{a\bar{b}}d\chi^a\wedge
d\chi^{\bar{b}}-K_{ia}d\phi^i\wedge
d\chi^a-K_{\bar{i}a}d\phi^{\bar{i}}\wedge d\chi^a)~,
\end{equation}
and we can see that the term is locally exact
\begin{equation}
L_{top}= -X^*\left ( d(K_ad\chi^a) \right )~. \label{toptermfull22}
\end{equation}
In other words, this is a total derivative term, but is not globally defined.

\section{Gerbes and the Total Derivative term}
\label{gerbes}

In this section we use gerbes (see Appendix for some basic facts)
  to elaborate on the geometrical meaning of the quasi-topological
  term\footnote{Related discussions of gerbes in the context of WZW models may be found in, e.g., \cite{Gawedzki:2002se} \cite{Schreiber:2005mi}. } (\ref{toptermfull22}).  
  We argue that although the   action is not well-defined, the path integral is.

  Before proceeding, it will be useful to compare two gauge choices for the $B$-field.
In a patch $U_\alpha$, 
we can choose the gauge 
$(B_\alpha) =(B_\alpha)_+$ in which the (1,1) part of $B$ with respect to $J_+$ vanishes, or the gauge 
 $(B_\alpha) =(B_\alpha)_-$  used in \cite{Gates:1984nk} in which the (1,1) part of $B$ with respect to $J_-$ vanishes, so that
 the  gauges for $B_\pm$ are
 \beq
  (B_\alpha)_\pm =   (B_\alpha)_\pm^{(2,0)} + (B_\alpha)_\pm^{(0,2)}~,
 \eeq
 with $B^{(1,1)}_\pm$ is zero.
 (Explicitly, $(B_\alpha)_+^{(2,0)}$ is the (2,0) part of $B$ with respect to $J_+$ and $(B_\alpha)_-^{(2,0)}$ is the (2,0) part of $B$ with respect to $J_-$.)
 These two gauge choices differ by a globally defined exact form
  \beq
  (B_\alpha)_+ - (B_\alpha)_-= 2d \Lambda_\alpha
 \eeq
 where $\Lambda_\alpha=\Lambda$ is a global 1-form.

As  $H$ is of  type $(2,1)+(1,2)$ with respect to both complex structures, it follows that
 \beq\label{HHH12002}
  H^{(2,1)}_\pm = d (B_\alpha)_\pm^{(2,0)}~.
  \eeq
 In the coordinate system used in previous sections, the explicit form of $(B_\alpha)_\pm^{(2,0)} $ is
$$
(B_\alpha)_-^{(2,0)} = K_{\bar{b}i}~ d\chi^{\bar{b}} \wedge d\phi^i~,~~~~~~~~~
(B_\alpha)_+^{(2,0)} = K_{ia} ~d\phi^i \wedge d\chi^a~.
$$
 It is possible to choose
  transition functions for the gerbe to be holomorphic with respect to both complex structures. It is thus natural to talk about bi-holomorphic gerbes.
Further details about the bi-holomorphic gerbe will be given in \cite{biholgerbe}.

 The term (\ref{toptermfull22}) is the pull-back of a locally exact form, which we denote $b$.
 Since the potential $K^{(\alpha)}$ is defined only locally  over a patch $U_\alpha$, this form is also defined
  only locally 
 \begin{equation}
b_{\alpha}=d(K^{(\alpha)}_a d\chi^{a})~,~~~~~~~b_\alpha \in \Omega^2(U_\alpha)~,~~~~
 K^{(\alpha)} \in C^\infty (U_\alpha)~.
\label{B expression}
\end{equation}
 Thus $\{b_\alpha\}$ is a collection of locally defined closed  complex forms on $M$.  These
  forms can be written as follows,
 \begin{equation}
 b_\alpha = (B_{\alpha})_-^{(0,2)} - (B_\alpha)_+^{(2,0)} + \frac{i}{2} \left ( \omega_- - \omega_+ \right )~.
\label{uglyformula}
\end{equation}

   The real and imaginary parts of $b_\alpha$ are
   \beq\label{balphacomplex}
  b_\alpha = \frac{i}{2} (F^+_\alpha + F^-_\alpha) + \frac{1}{2} (B_\alpha)_- - \frac{1}{2}(B_\alpha)_+~
 = \frac{i}{2} (F^+_\alpha + F^-_\alpha) +d \Lambda~,
  \eeq
  where $F_\alpha^\pm \in \Omega^2 (U_\alpha)$ are the real closed two-forms on $U_\alpha$
   defined as follows
  \beq
   F_\alpha^\pm = i \left ( (B_\alpha)_\pm^{2,0} - (B_\alpha)_\pm^{0,2} \right ) \mp \omega_\pm ~.
\label{fis}
 \eeq
   The property $dF^\pm_\alpha=0$ is a consequence of the conditions (\ref{integrabilityforGKS})
     and (\ref{HHH12002}).  Thus, in a way the forms $F^\pm_\alpha$ are the local analogue  of the K\"ahler 
      form in the K\"ahler geometry\footnote{In  K\"ahler geometry 
     $ \nabla J = 0$ is equivalent to $d\omega = 0$, while in our case $ \nabla^{\pm} J_{\pm} = 0$ is locally equivalent to $d F_{\pm} = 0.$}. 
     The real part $d\Lambda$ is an exact form and does not contribute to the integral, so that the quasi-topological term in the action is
   \beq\label{stop}
  S_{top} = i  \int\limits_{\Sigma_2} X^*(F) =i \int\limits_{X_*(\Sigma_2)}   F~,
    \eeq    
    where
      \beq\label{stop}
   F_\alpha = \frac{1}{2} (F^+_\alpha + F^-_\alpha) ~.
    \eeq

  Next, we check that 
    this   term reduces to the familiar topological terms in the standard A- and B-models.
   In the standard K\"ahler case when $J_-= -J_+$ and $B$ is a globally
    defined closed two-form,
      $F$ is the complexified K\"ahler class
      \beq
       iF =  B + i \omega~.
      \eeq
      Then $S_{top}$ is the topological term for the A-model as expected. 
    In the standard K\"ahler case with $B=0$ and   $J_-= +J_+$,
    $F=0$ and there is no topological term, as expected for the B-model.
    It is interesting that in this case, if we introduce a
    $B$ which is a globally
    defined closed two-form,
    then
    $$iF=-\left ( B^{2,0} - B^{0,2} \right )~,
  $$
  so that
  $$
  S_{top} = - \int \left ( B^{2,0} - B^{0,2} \right )~,
$$
is the integral of a global 2-form and is well-defined despite the absence of an \lq $i$'.

We now return to the general case in which $B$ is a gerbe connection.
   By itself the term $S_{top}$ is not well-defined. However we can make sense of $S_{top}$
    by exponentiating and interpreting this as a holonomy of a flat gerbe.
     Let us briefly recall the case of Wilson loops and flat connections on line bundles. For a line bundle with a flat connection,  the connection $A$ is a collection of locally defined closed 1-forms, with suitable transition functions.
For a line bundle with connection $A$, the holonomy operator for a curve $\gamma$ (a Wilson loop) is
\begin{equation}
W_A=\exp \left( {i\oint\limits_\gamma A} \right)~,
\end{equation}
and if $A$ is flat this  depends only on the homology class of $\gamma$. Then  for flat line bundles, $W_A$
 defines a map
$$W_A ~:~H_1(M) ~\rightarrow~ S^1~,$$
 so that the map is an element of $H^1(M,U(1))$.
 
 The same idea works for a real  flat gerbe.
        For a gerbe with a connection $b$ there exists
 a holonomy operator defined for any 2-cycle $\Sigma$
 \begin{equation}
W_b=\exp \left( {i\oint\limits_{\Sigma} b} \right)~,
\end{equation}
and if  the connection is flat, $db=0$, 
 this  depends only on the homology class of $\Sigma $.
Then for flat connections, $W_b$ 
defines a map
$$W_b~:~ H_2(M)~\rightarrow~S^1~,$$
which is then an element of $H^2(M,U(1))$.
 The flat gerbe is a collection of locally defined closed 2-forms, with suitable transition functions. The operator $W_b$  depends only on the  homology of $\Sigma $ and the connection $b$.

 Now we would like to apply the idea of a flat gerbe to our quasi-topological term $S_{top}$.
   Upon exponentiating this term and interpreting as a holonomy operator
  \begin{equation}
   \exp \left( i
   { \int\limits_{X_*(\Sigma_2)}  F} \right)~:~H_2(M)~\rightarrow~ S^1~,
  \label{finaltermsss}
  \end{equation}
     we arrive at a term which depends only on the homology class of $X_*(\Sigma_2)$.
        Thus  finally we conclude that, independent of the gauge,
      the topological term should be understood through the holonomy  of a real flat 
      gerbe connection\footnote{For notation, see \cite{Belov:2007qj}.},
      \begin{equation}
   \exp \left( i
   { \int\limits_{\Sigma}  F} \right)=W_F= Hol \left ( F \right )~,
  \label{finaltermsss122}
  \end{equation}
   where $\Sigma = X_*(\Sigma_2)$. 
   
  \section{Discussion}
\label{end}

We have shown that the path integral can be written as a weighted sum for the B-twisted BiLP models, i.e., for twisted sigma models involving chiral and twisted chiral fields
(and thus H-flux). One of the terms in the action corresponds to a flat gerbe connection and has an interpretation as a quasi-topological term in the quantized theory. Its exponential is a Wilson surface, a generalization of a Wilson line. We have further seen that the localization of the model is on holomorphic twisted chiral maps and constant chiral ones.

A natural question is how to extend the discussion to the full Generalized K\"ahler Geometry, i.e., to include semi-chiral fields in the model. Indeed we believe that our
 result will extend to the general twisted  ${\cal N}=(2,2)$ model, including the semichiral fields.
 For the general case in which $J_+$ and $J_-$ do not commute, 
 then the  $F^\pm$ defined by (\ref{fis}) still satisfy $dF^\pm=0$  and so are flat gerbe connections.
 We conjecture that the twisted theory continues to be weighted by 
 the holonomy of the gerbe connection $F = {1\over 2} (F^+ +F^-)$
 in this general case, generalising the exponential of the degree.
  There are a number of important issues to be addressed regarding how to deal with  the 
   the ${\cal N}=(2,2)$ Euclidean model. We plan to resolve these and related problems in the forthcoming publications. 
\bigskip

{\bf Acknowledgement:}  We are grateful to Martin Ro\v{c}ek for discussions, and to Ezra Getzler for help with gerbe holonomies. 
The research of U.L. was supported
by EU grant (Superstring theory) MRTN-2004-512194 and by VR grant 621-2006-3365.
 The research of R.v.U. was supported by
Czech ministry of education contract No.~MSM0021622409. The research
of M.Z. was supported by VR-grant 621-2004-3177. The research of
L.M.d.S. was supported by FCT grant SFRH/BD/10877/2002.

\appendix
\section{Gerbe holonomy}
\label{appendix}

In this Appendix we briefly review the notion of holonomy for   line bundles 
 and   gerbes. For more details the reader may consult \cite{Belov:2007qj, hitchingerbe, chatterjee, mackaay}. 
 
 Consider a smooth manifold $M$ with an open covering $\{ U_\alpha\}$ where all open sets and intersections are contractible. 
   The line bundle can be thought of as a
set of transition functions
$$  g_{\alpha\beta}~:~U_\alpha\cap U_\beta~\rightarrow~S^1~,$$
which satisfy $g_{\alpha\beta} = g^{-1}_{\beta\alpha}$ and the cocycle condition on $U_\alpha\cap U_\beta \cap U_\gamma$
$$   g_{\alpha\beta} g_{\beta\gamma} g_{\gamma\alpha} = 1~.$$
  The connection on the line bundle can be defined as a collection of one-forms $A_\alpha \in \Omega^1(U_\alpha)$ such that on the double intersections $U_\alpha \cap U_\beta$
$$      iA_\alpha - i A_\beta = g_{\alpha\beta}^{-1} d g_{\alpha\beta}~.$$ 
 Since on $U_\alpha\cap U_\beta$ $dA_\alpha = dA_\beta$ we can define a curvature two 
  form $\omega$ on $M$ such that $\omega = dA_\alpha$ on $U_\alpha$. It can be shown that $\omega$ defines an integral cohomology class, 
   $\omega/2\pi \in H^2(M, \mathbb{Z})$.  
 
 For any  loop $\gamma$ in $M$,  the holonomy is defined as follows. First (assuming a suitably fine open cover) the loop $\gamma$  is divided into segments $\gamma_\alpha$ such that each  $\gamma_\alpha$ is in $U_\alpha$ and the point (if any) at which 
 $\gamma_\alpha$ and $\gamma_\beta$ join is denoted $\gamma_{\alpha\beta}$.
  Then the holonomy of $A$ on the curve $\gamma$ is
 $$ Hol (A, \gamma) = \exp \left ( i \sum\limits_{\gamma_\alpha} \int\limits_{\gamma_\alpha} A_\alpha
  +  i \sum\limits_{\gamma_{\alpha\beta}} \log g_{\alpha\beta}(\gamma_{\alpha\beta})\right )~,$$
  and it can be shown that it does not depend on a particular choice of the partition $\gamma$ into $\{ \gamma_\alpha, \gamma_{\alpha\beta}\}$.  If $\omega=0$, then there is flat connection on a line bundle. 
      In this case the holonomy $Hol(A, \gamma)$ depends only on the homology class of $\gamma$.
 
  A gerbe is a higher generalization of a line bundle. A gerbe can be defined as a set of transition 
   functions on threefold intersections
$$   g_{\alpha\beta\gamma}~:~ U_\alpha \cap U_\beta \cap U_\gamma ~\rightarrow ~S^1~,$$
satisfying 
$$    g_{\alpha\beta\gamma} = g^{-1}_{\beta\alpha\gamma} = g^{-1}_{\alpha\gamma\beta} =
     g^{-1}_{\gamma\beta\alpha}~,~~~~~~  
     g_{\beta\gamma\delta} g_{\delta\gamma\alpha}g_{\alpha\beta\delta} 
     g_{\beta\alpha\gamma}=1~,$$
where the last condition is understood on $U_\alpha\cap U_\beta\cap U_\gamma \cap U_\delta$.
   A connection on a gerbe is defined as a collection of one-forms and two-forms
    $\{ A_{\alpha\beta}, B_\alpha\}$ such that  $A_{\alpha\beta} \in \Omega^1(U_\alpha\cap U_\beta)$ and  and $B_{\alpha} \in \Omega^2 (U_\alpha)$ with the relations
$$   i A_{\alpha\beta} + i A_{\beta\gamma} + i A_{\gamma\alpha} = g^{-1}_{\alpha\beta\gamma}
   d g_{\alpha\beta\gamma}~,$$
    on the triple intersections $U_\alpha \cap U_\beta \cap U_\gamma$ and 
$$B_\alpha - B_\beta = d A_{\alpha \beta}~,$$
 on the double intersection $U_\alpha\cap U_\beta$. 
  Since  $dB_\alpha = dB_\beta$ on $U_\alpha\cap U_\beta$, one  can define a curvature three 
  form $H$ on $M$ such that  $H = dB_\alpha$ on $U_\alpha$. It can be shown that $H$ defines an integral cohomology class,  
   $H/2\pi \in H^3(M, \mathbb{Z})$.  
   
   For any closed 2-surface $\Sigma$ in $M$,  the holonomy of a gerbe with connection is defined
    as follows. We choose an open cover $U_\alpha$ of $M$ and a simplicial decomposition of $\Sigma$
    into 2-simplices $\Sigma_{\alpha}$ such that 
$\Sigma_{\alpha}$  is in $U_\alpha$
and   if $\Sigma_\alpha$ and $\Sigma_\beta$ have a common edge, that 1-simplex is labelled $\Sigma_{\alpha\beta}$. If the three 1-simplices $\Sigma_{\alpha\beta}$, $\Sigma_{\beta\gamma}$, $\Sigma_{\gamma\beta}$ intersect in a point, it is labelled $\Sigma_{\alpha\beta\gamma}$.
The gerbe holonomy is then
     $$ Hol (B, A, \Sigma) = \exp \left (i \sum\limits_{\Sigma_\alpha}~ \int\limits_{\Sigma_\alpha} B_\alpha
     +      i \sum\limits_{\Sigma_{\alpha\beta}}~ \int\limits_{\Sigma_{\alpha\beta}} A_{\alpha\beta}
  +  i \sum\limits_{\Sigma_{\alpha\beta\gamma}} \log g_{\alpha\beta\gamma}(\Sigma_{\alpha\beta\gamma})\right )~,$$
 and one can prove that this does not depend on the particular choice of open cover of $M$ or of simplicial decomposition of $\Sigma$ into 
     $\{ \Sigma_\alpha, \Sigma_{\alpha\beta}, \Sigma_{\alpha\beta\gamma}\}$. 
  If $H=0$ the gerbe is called flat. For a flat gerbe the holonomy depends only on the homology 
   class of $\Sigma$.

\eject

\end{document}